\pdfoutput=1 

\documentclass[sigconf]{acmart}

\usepackage{dirtytalk}
\usepackage{siunitx}
\usepackage[inline]{enumitem}
\newcommand{\co}{compiler}
\newcommand{\wa}{warning}
\newcommand{\cw}{compiler warning}
\newcommand{\clang}{{C}lang}
\newcommand{\gcc}{\textsc{GCC}}
\newcommand{\msvc}{\textsc{MSVC}}

\acmConference[ICSE 2022]{The 44th International Conference on Software Engineering}{May 21–29, 2022}{Pittsburgh, PA, USA}

\title{The Unexplored Terrain of Compiler Warnings}

\begin{abstract}
The authors' industry experiences suggest that compiler warnings,
a lightweight version of program analysis,
are valuable early bug detection tools.
Significant costs are associated with patches and security bulletins
for issues that could have been avoided if compiler warnings were
addressed.
Yet, the industry's attitude towards compiler warnings is mixed.
Practices range from silencing all compiler warnings to having a
zero-tolerance policy as to any warnings.
Current published data indicates that addressing compiler warnings
early is beneficial.
However, support for this value theory stems from grey literature
or is anecdotal.
Additional focused research is needed to truly assess the cost-benefit
of addressing warnings.
\end{abstract}

\copyrightyear{2022}
\acmYear{2022}
\setcopyright{rightsretained}
\acmConference[ICSE-SEIP '22]{44nd International Conference on Software Engineering: Software Engineering in Practice}{May 21--29, 2022}{Pittsburgh, PA, USA}
\acmBooktitle{44nd International Conference on Software Engineering: Software Engineering in Practice (ICSE-SEIP '22), May 21--29, 2022, Pittsburgh, PA, USA}
\acmDOI{10.1145/3510457.3513057}
\acmISBN{978-1-4503-9226-6/22/05}

\begin{document}

\author[1]{Gunnar Kudrjavets}
\orcid{0000-0003-3730-4692}
\affiliation{
   \institution{University of Groningen}
   \city{Groningen}
   \country{Netherlands}}
\email{g.kudrjavets@rug.nl}

\author[2]{Aditya Kumar}
\affiliation{
    \institution{Snap, Inc.}
    \streetaddress{2772 Donald Douglas Loop N}
    \city{Santa Monica}
    \state{CA}
    \country{USA}
    \postcode{90405}}
\email{adityak@snap.com}

\author[3]{Nachiappan Nagappan}
\affiliation{
    \institution{Microsoft Research$^{\ast}$}
    \streetaddress{One Microsoft Way}
    \city{Redmond}
    \state{WA}
    \country{USA}
    \postcode{98052}}
\email{nnagappan@acm.org}
\thanks{$^\ast$ This work was initiated while Nachiappan Nagappan was with Microsoft Research. He is currently with Facebook, Inc.}

\author[4]{Ayushi Rastogi}
\affiliation{%
   \institution{University of Groningen}
   \city{Groningen}
   \country{Netherlands}}
\email{a.rastogi@rug.nl}



\maketitle

\section{Opportunity and motivation}

One of the earliest stages of software development during which bugs can
be detected is when new code changes are compiled.
A compiler can flag \emph{potential} issues found in the code.
The cost of fixing a software defect increases significantly during the
later phases of the development cycle~\cite{tassey_2002}.
It is optimal to correct problems as early as possible because
even minor bugs can result in catastrophic consequences~\cite{zhivich_2009}.
Classical memory safety related bugs, common to programming languages such as
C and C++ (e.g., a double-free), are the reasons behind approximately 70\% of
security updates Microsoft issues each year~\cite{matt_trends_2019}.
For Microsoft, the cost of fixing a bug resulting in
a security bulletin is approximately \SI{100000}[\$]{}~\cite[p.~11]{howard_2002}.
Modern compilers (e.g., \clang, \gcc, and \msvc) can detect typical programming
mistakes such as integer overflows or underflows, out-of-bounds errors,
memory management problems, etc., either during the compilation phase or via
using runtime sanitizers.
Acting on \cw s enables engineers to fix defects early and prevent the
cascading set of failures the bugs would have otherwise caused.
Despite the potential benefits of heeding \cw s,
industry attitudes are mixed.
For Linux kernel development, it took thirty years to start treating \wa s as errors~\cite{linux_2021}.
Google takes a somewhat contrarian approach by aiming to never issue
\cw s because they find that developers ignore them~\cite[p.~427]{winters_2020}.
The \wa s are either enabled as errors or never shown in the compiler output.

In our experience, demonstrating the value of fixing \cw s or
changing organizational culture to treat \cw s as a first-class
defect prevention tool is challenging.
Often, changes in attitude and engineering processes only take place
after damaging events (e.g., critical services becoming inaccessible,
irrecoverable data loss, zero-day exploits) have already manifested.

\section{Existing evidence and guidance}

Existing research and empirical data about the benefits of fixing \cw s is minimal.
Most of the data comes from grey literature related to writing
secure code or anecdotal knowledge passed down from experienced practitioners.

Microsoft practices recommend using the highest level of \wa s to inspect
code for potential security vulnerabilities and compile \say{cleanly}
without any errors or warnings~\cite{howard_2006,howard_sdl_2006}.
The downside of not fixing the \wa s is articulated in a case study on a
large code base where integer-related \wa s had been disabled.
Analysis reveals that about $20\%$ of the hidden \wa s contain
potentially exploitable conditions~\cite{howard_2007}.
A post-mortem analysis from Facebook finds that enabling all compiler
warnings as errors %
reveals issues such as memory leaks, infinite recursion,
and catastrophic bugs where a compiler would \say{optimize}
away critical functions~\cite{kumar_2019}.
Reducing the attack surface and finding opportunities to
clean up code during the maintenance phase is another
suggested application for utilizing \cw s~\cite{pearse_1995}.

We can find only one paper investigating the correlation
between \cw s and defects~\cite{moser_2007}.
The study finds experimental evidence that \say{[a] large number of
compiler warnings of a source file is an indicator that the file
contains also an above-average number of defects.}
The conclusion is based on a limited amount of data
and uses a version of \gcc\ from 2006.
Given the advances in compiler technology during the last 16 years and
the size of industrial code bases (e.g., in 2017 the Windows code
base contained $3.5$ million files) we need additional studies utilizing the latest versions of compiler toolsets and larger projects~\cite{harry_2017}.
The remaining discoverable research related to \cw s is focused on their correctness,
readability, and validity~\cite{barik_2018,sun_2016}.

\section{Observations from industry}

The trend we observe is that \emph{an engineer's experience and seniority
are directly related to his or her attitude towards fixing \wa s}.
The more experience with the cost and consequences of basic
programming errors the engineers have,
the more appreciative they are of ensuring
the correctness of the code as early as possible.

From a technical point of view, \emph{we rarely observe projects treating
\wa s as errors and triggering build breaks as a result}.
Turning on \emph{all possible warnings}
is mainly done by engineers developing \co s themselves.
Very few projects in industry have a zero-tolerance policy towards
the presence of \cw s.
A rare example is safety-critical code, e.g., software developed by \textsc{NASA}~\cite{holzmann_2006}.
We have not been able to find any public data regarding standards
related to \cw s
in other companies producing safety-critical software, e.g.,
Airbus, Boeing, Tesla, etc.

A variety of reasons contribute to \cw s either not
being fixed or deprioritized.
The main reason is the \emph{lack of empirical evidence} to show either
correlation or causal relationship between fixing
\cw s and decrease in defect density.
Another key reason is the \emph{cost of adapting stricter \cw\ levels to legacy code}.
Techniques such as treating warnings as errors are time-consuming
to implement unless projects established this policy from
the very beginning.
We cannot discount the \emph{impact on an engineer's career} as well.
The lack of external motivation to fix the \cw s is often caused
by the fact that preemptively fixing \cw s does not get rewarded
as well as the post hoc activity associated with debugging and bug fixing.
The \emph{repetitive nature of fixing the \cw s} is another factor making
long-term code quality improvement initiatives unpopular.
The number of warnings to be analyzed may reach into hundreds, thousands or
even tens of thousands depending on the size of the code base.
A key reason related to engineers not willing to fix
\cw s is %
\emph{distrust in the validity of the warnings} due to past
experiences with false positives.
This belief can be
summarized as \say{if warnings would indicate \emph{real problems}, then
they would be errors instead}.

\section{Future research directions}

We recommend that researchers partner with practitioners
working on open- and closed-source software to focus on
following topics:

\begin{enumerate}
    \item \emph{Explore the current state}.
    What are the default warning levels, attitudes and sets of beliefs
    toward fixing the warnings?
    Are they influenced by software's technical abstraction level?
    \item \emph{Investigate the relationship (or lack thereof) between \cw s
    and defects, team productivity, and product risk}.
    \item \emph{Establish baseline metrics related to \cw s}.
    For example, \wa s per file, per \textsc{KLOC}, change in the ratio of \wa s with
    the application of stricter levels of compilation,
    number of suppressed \wa s per \textsc{KLOC}?
    \item \emph{Rank \wa\ categories  according to their precision and recall}.
    Propose a recommended set of \wa s per compiler.
    \item \emph{Conduct case studies about projects having zero-tolerance policy
towards \wa s}.
    Is the approach cost-effective outside the scope of safety-critical software?
    \item \emph{Evaluate the economics (e.g., negative impact) of fixing \wa s}.
    {SQLite} development team finds that
    \say{[m]ore bugs have been introduced into SQLite while trying to get
    it to compile without warnings than have been found by static analysis}~\cite{sqlite_2021}.
    \item \emph{Variation between programming languages}.
    How similar are or should be \wa s for low-level (e.g., {C}),
    functional (e.g., {OC}aml),
    or scripting (e.g., {R}uby) languages?
\end{enumerate}

\bibliographystyle{ACM-Reference-Format}
\bibliography{warnings}

\end{document}